\documentclass[12pt]{article}
\usepackage{epsfig}

\def\appendix#1{
  \addtocounter{section}{1}
 \setcounter{equation}{0}
  \renewcommand{\thesection}{\Alph{section}}
 \section*{Appendix \thesection\protect\indent \parbox[t]{11.715cm} {#1}}
  \addcontentsline{toc}{section}{Appendix \thesection\ \ \ #1}
  }

\newcommand{\newsection}{
\setcounter{equation}{0}
\section}

\def\bea{\begin{eqnarray}}
\def\eea{\end{eqnarray}}
\def\be{\begin{equation}}
\def\ee{\end{equation}}

\newcommand{\rf}[1]{(\ref{#1})}
\newcommand{\non}{\nonumber \\*}
\def\al{\alpha}
\def\kq{\kappa_Q}
\def\kb{\kappa_B}
\def\ke{\kappa_{L^e}}\def\km{\kappa_{L^\mu}}\def\kt{\kappa_{L^\tau}}

\begin{document}
\begin{titlepage}
\begin{flushright}
ITEP--TH--43/00\\
\end{flushright}
\vspace{1.5cm}

\begin{center}
{\LARGE Lepton Asymmetry of the Universe} 
\\[.5cm]
{\LARGE and Charged Quark-Gluon Plasma}\\
\vspace{1.9cm}
{\large K.~Zarembo}\\
\vspace{24pt}
{\it Department of Physics and Astronomy}
\\{\it and Pacific Institute for the Mathematical Sciences}
\\{\it University of British Columbia}
\\ {\it 6224 Agricultural Road, Vancouver, B.C. Canada V6T 1Z1} 
\\ \vskip .2 cm
and\\ \vskip .2cm
{\it Institute of Theoretical and Experimental Physics,}
\\ {\it B. Cheremushkinskaya 25, 117259 Moscow, Russia} \\ \vskip .5 cm
E-mail: {\tt zarembo@physics.ubc.ca}
\end{center}
\vskip 2 cm
\begin{abstract}
The lepton asymmetry of order one is not excluded experimentally and,
if present,
can lead to interesting phenomena in the early Universe. 
It is shown that, when
the temperature is above the quark-hadron transition,
the lepton asymmetry induces chemical potentials for electric
charge and for baryon number, and makes
the quark-gluon plasma electrically charged.
\end{abstract}

\end{titlepage}
\setcounter{page}{2}
\newsection{Introduction}

The observational data do not severely restrict lepton asymmetry of 
the Universe which, if present, resides almost entirely
in the neutrino sector. In particular, 
the chemical potentials and the temperature of relic neutrinos
can be comparable in magnitude. The nucleosynthesis
bounds on
the neutrino degeneracy parameters, $\xi_{\nu_f}=\mu_{\nu_f}/T_\nu$,
are $-0.06<\xi_{\nu_e}< 1.1$, $|\xi_{\nu_{\mu,\tau}}|< 6.9$
\cite{Kan92}. 
Current measurements of the
cosmic microwave background anisotropy put an upper limit of $3-5$ on
degeneracy parameters for all neutrino species \cite{Kin99,Les99}.
 
The large lepton asymmetry could have profound consequences
in the early history of the Universe, when the temperature was
of order of the electroweak scale.
The lepton charge alters the pattern of electroweak symmetry breaking,
because the lepton number density forces
the Higgs field to Bose condense  and therefore can impede
restoration of electroweak symmetry at high temperature 
\cite{Lin76,Hab82,Rio97}.
 If the lepton number density is 
sufficiently large, the Universe has always been in the symmetry-broken phase
and there was no electroweak phase transition \cite{Lan82}. This
observation provides a possible resolution
of the monopole and the domain wall
problems \cite{Baj98}. 

Although it seems 
natural to assume that the lepton and the baryon asymmetries
of the Universe are of the same order of magnitude, 
mechanisms that could generate primordial lepton asymmetry of order
one without generating large baryon asymmetry are known
\cite{Har81,Liu94,Cas99,Mar99}, and degeneracy of
relic neutrinos remains at least a logical possibility.
The assumption of large neutrino degeneracy 
was invoked, in particular, 
to explain an origin of the ultra-high energy
cosmic rays \cite{Gel99}.

In this paper I will study the consequences of 
 the large lepton asymmetry at 
temperatures above the quark-hadron transition but below the
electroweak scale. It appears that,
when quarks are deconfined, the lepton
asymmetry indirectly affects hadronic (strongly interacting) sector
in such a way that all light quark species acquire large chemical potentials.
A deviation of 
the lepton number density from zero and an overall electric neutrality
cause an asymmetric distribution of the electric charge between
the lepton and the hadron sectors. 
As a consequence, the quark-gluon plasma is electrically charged
when 
lepton asymmetry is of order one. 

\newsection{Chemical potentials}

There are five charges that are conserved 
or almost conserved when the temperature
drops below the electroweak scale: 
electric charge, baryon number and lepton numbers for
each of the three generations of fermions. Accordingly, one can 
introduce five chemical
potentials to describe a state with non-zero densities
of these charges. On average, two of the 
conserved charge densities are zero, because
the Universe must be
electrically neural and the baryon asymmetry of the Universe is
negligibly small. 
But densities of lepton numbers can potentially be large enough to 
require chemical potentials comparable to the temperature.

The chemical potential for each of the particle species 
is a linear combination of the chemical potentials for conserved
charges:
\be\label{mus}
\mu_i=\sum_\al q_i^\al\mu_\al.
\ee
Here $q_i^\al=(Q_i,B_i,L_i^e,L_i^\mu,L_i^\tau)$ are charges of $i$th
species. In assumption of large lepton asymmetry, chemical potentials
for lepton numbers are non-zero at present and cause asymmetries
in equilibrium distributions of the relic neutrinos. When the temperature
was higher than the mass of the electron, the lepton number chemical
potential alone  would lead to an asymmetric distribution of electrons and
would induce a non-zero density of electric charge. The electric
neutrality thus requires to set $\mu_Q=\mu_{L^e}$, which
makes the chemical potential for electrons equal to zero.

The situation drastically changes when 
quarks are deconfined. Light quarks can be considered essentially
massless at temperatures above the quark-hadron transition and,
since they carry both the electric
charge and the baryon number, 
non-zero $\mu_Q$ would induce large baryon asymmetry unless
$\mu_B$ is also non-zero. The conditions
of electric neutrality and baryon symmetry then become more complicated and
 all chemical potentials generically become of the same
order of magnitude.

I will consider temperatures of order of few Gev, when the temperature 
suppresses strong interactions due to the asymptotic freedom,
 and the ideal gas
approximation for the
quark-gluon plasma is more or less accurate.
The active species 
at such temperatures are neutrinos, electrons, muons, photons,
gluons  and u, d and s quarks. Weak processes at the QCD epoch are rapid
enough to maintain thermal equilibrium, and 
I will also assume that strong 
interactions do not considerably distort equilibrium
distributions and will neglect masses of all active particle
species.

The excess of particles over anti-particles per unit volume for two-component
non-interacting fermions with chemical potential $\mu$ at temperature 
$T$ is 
\be
n_+-n_-=\frac{T^3}{6}\left(\kappa+\frac{\kappa^3}{\pi^2}\right).
\ee
The degeneracy parameters at the QCD epoch will be denoted by $\kappa$: 
$\kappa=\mu/T$, to distinguish them from the degeneracy parameters now, which
are denoted by $\xi$. With the help of this equality,
the electric neutrality, absence of
considerable baryon number and requirement that there is a given
lepton asymmetry yield a set of five equations for five
chemical potentials:
\bea\label{main}
\frac{T^3}{6}\sum_i Q_i\left(\kappa_i+\frac{\kappa_i^3}{\pi^2}\right)
=&0&\non
\frac{T^3}{6}\sum_i B_i\left(\kappa_i+\frac{\kappa_i^3}{\pi^2}\right)
&=&0\non
\frac{T^3}{6}\sum_i L^f_i\left(\kappa_i+\frac{\kappa_i^3}{\pi^2}\right)
&=&n_{L^f}-n_{\bar{L}^f},
\eea
where $n_{L^f}$ ($n_{\bar{L}^f}$) are number densities of leptons 
(anti-leptons) of flavor $f$.

Though the lepton numbers are not conserved exactly, and some
lepton charge could be produced by
neutrino oscillations during nucleosynthesis
\cite{Foo96,Shi96,Dib00}, the asymmetry generated in this
way is generically much smaller than one \cite{Dol99,Dol00}. 
So, if large density of lepton number was 
present at the QCD epoch, later it was just diluted by the expansion of the
Universe. Since the entropy density,
\be
s=\frac{P+\rho-\mu(n_+-n_-)}{T},
\ee
 scales in the same way\footnote{This 
assumption implies an entropy conservation and,
in particular, an absence of considerable entropy production at
the QCD phase transition.}, the ratio of the lepton number density
to the entropy is time-independent:
\be
n_{L^f}-n_{\bar{L}^f}=s\left(\frac{n_{L^f}-n_{\bar{L}^f}}{s}\right)_{\rm now}. 
\ee
Taking into account that the entropy
of a gas of non-interacting two-component fermions is
\be
s_{\rm ferm}=T^3\left(\frac{7\pi^2}{90}
+\frac{\kappa^2}{6}\right),
\ee
and that each bosonic degree of freedom contributes $2\pi^2T^3/45$, we find:
\be
s=T^3\left(\frac{247\pi^2}{90}+\sum_i\frac{\kappa^2_i}{6}\right)
\ee
at the QCD epoch.
The entropy at present is carried by photons and by neutrinos:
\be
s_{\rm now}=\frac{4\pi^2}{45}\,T_\gamma^3+T^3_\nu\sum_f\left(\frac{7\pi^2}{90}
+\frac{\xi_{\nu_f}^2}{6}\right)=T_\nu^3\left(\frac{43\pi^2}{90}+
\sum_f\frac{\xi_{\nu_f}^2}{6}\right),
\ee
where the relation 
$T_\gamma^3=11T^3_\nu/4$
 was used in the last equality. The present value
of the lepton asymmetry depends on
the neutrino chemical potentials as follows:
\be
\left(n_{L^f}-n_{\bar{L}^f}\right)_{\rm now} 
=\frac{T^3_\nu}{6}\left(\xi_{\nu_f}+\frac{\xi_{\nu_f}^3}{\pi^2}\right).
\ee
Finally, we get:
\be\label{assym}
n_{L^f}-n_{\bar{L}^f}=\frac{T^3}{6}\left(\frac{247\pi^2}{15}+\sum_i\kappa_i^2
\right)
\frac{\xi_{\nu_f}+\frac{\xi_{\nu_f}^3}{\pi^2}}{\frac{43\pi^2}{15}
+\xi_{\nu_e}^2+\xi_{\nu_\mu}^2
+\xi_{\nu_\tau}^2}\,.
\ee

Plugging quantum numbers\footnote{I
normalize the
baryon charge so that quarks have baryon number 
$1/3$.} of active species in eq.~\rf{mus},
using eq.~\rf{assym} and
the identities for electric charges of light quarks:
\be
\sum_{u,d,s}Q_i=0,~~~\sum_{u,d,s}Q_i^2=\frac23,~~~
\sum_{u,d,s}Q_i^3=\frac29,~~~\sum_{u,d,s}Q_i^4=\frac29,
\ee
we get a system of five equations for dimensionless degeneracy
parameters $\kappa_\alpha=\mu_\alpha/T$:
\bea\label{more}
&&\frac83\,\kq-\ke-\km+\frac{2}{9\pi^2}(\kq^3+\kq^2\kb+\kq\kb^2)
-\frac{1}{\pi^2}(\ke-\kq)^3-\frac{1}{\pi^2}(\km-\kq)^3=0
\non
&&\kappa_B^3+(6\kappa_Q^2+9\pi^2)\kappa_B+2\kappa_Q^3=0
\non
&&2(\ke-\kq)+\frac{2}{\pi^2}(\ke-\kq)^3+\ke+\frac{1}{\pi^2}\,\ke^3
=\eta_e
\non
&&2(\km-\kq)+\frac{2}{\pi^2}(\km-\kq)^3+\km+\frac{1}{\pi^2}\,\km^3
=\eta_\mu
\non          
&&\kt+\frac{1}{\pi^2}\,\kt^3
=\eta_\tau,
\eea
where
\bea
\eta_f&=&\left[\frac{247\pi^2}{15}+\frac43\,\kq^2+\frac23\kb^2
+2(\ke-\kq)^2+2(\km-\kq)^2+\ke^2+\km^2+\kt^2\right]
\non
&&\times
\frac{\xi_{\nu_f}+\frac{\xi_{\nu_f}^3}{\pi^2}}{\frac{43\pi^2}{15}
+\xi_{\nu_e}^2+\xi_{\nu_\mu}^2
+\xi_{\nu_\tau}^2}\,.
\eea

The system of equations \rf{more} allows to expresses the degeneracy
parameters at the time in the history of the Universe when the temperature
was of order of few Gev in terms of neutrino degeneracy parameters now.
This system, 
upon closer look, does not contain any large numerical coefficients.
So, if $\xi_{\nu_f}$ are of order one, all $\kappa_\al$, 
including $\kb$ and $\kq$, will also be of order one. 
For instance, in the linear approximation,
$\kb=0$ and $\kq=1.44(\xi_{\nu_e}+\xi_{\nu_f})$.

\begin{figure}[ht]
\hspace*{5cm}
\epsfxsize=7cm
\epsfbox{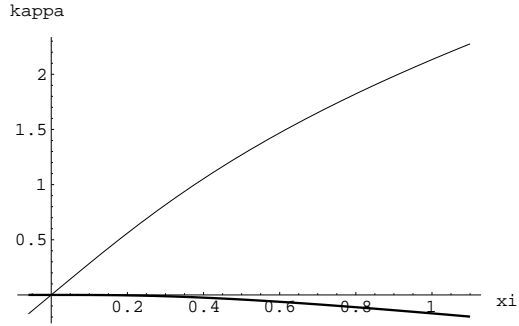}
\caption[x]{The electric charge
chemical potential and the baryon number chemical potential
(bold line) at the QCD epoch, $\kappa=\mu/T$, as the
functions of neutrino degeneracy parameters at present: 
$\xi_e=\xi_\mu=\xi_\tau\equiv\xi$.}
\label{dem}
\end{figure}

\begin{figure}[ht]
\hspace*{5cm}
\epsfxsize=7cm
\epsfbox{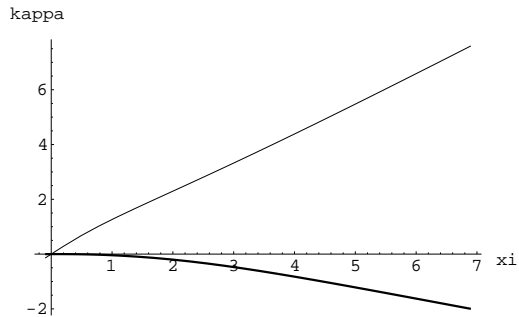}
\caption[x]{The same as in Fig.~\protect\ref{dem}, but for
$\xi_e=0$,
$\xi_\tau=0$, $\xi_\mu\equiv\xi$.}
\label{elec}
\end{figure}

I have solved the system of equations for the degeneracy parameters 
numerically in two representative cases. In the first case,
neutrino degeneracies were chosen 'democratically' -- equal for
 all three neutrino species: $\xi_e=\xi_\mu=\xi_\tau\equiv\xi$. The 
 dependence of
the baryon chemical potential and the chemical potential for electric charge
at the QCD epoch
on $\xi$ are shown in Fig.~\ref{dem}.
Fig.~\ref{elec} shows the same quantities in the second case,
when I took $\xi_e=0$ and
$\xi_\tau=0$.
  
\newsection{Discussion}

The main result of this paper is that
an overall charge neutrality, equilibrium with respect to the weak 
interactions and a non-zero density of the lepton number imply that
the quark-gluon sector of the primordial plasma in the early
Universe carries large (comparable to the entropy)
electric charge, which compensates the opposite charge in the lepton sector.
It is necessary to mention that, in
explicit calculations of the chemical potentials carried out in the 
previous section, the quark-gluon plasma
was treated as an ideal gas, which is an accurate approximation
 only at sufficiently high temperature.
In practice, the ideal gas approximation should 
work well only for extensive quantities (entropy, charge density, etc.)
and at temperature of order of few Gev or higher. 
The most interesting phenomena happen at lower temperatures when the ideal
gas approximation definitely breaks down. Still, the qualitative
conclusion that the quark-gluon plasma is charged in the presence of
lepton asymmetry should remain unchanged, because it relies solely on
the fact that quarks carry electric charges and
 can be thermally excited above the QCD
phase transition.

The physics of charged quark-gluon plasma can differ considerably from
that of the better studied neutral plasma.
The charge density potentially can affect
the fate of various topological defects \cite{Nag99,Koz99} and can
change
the nature of the quark-hadron transition.
At zero chemical potentials, the quark-gluon and the hadron phases are
likely to be  smoothly connected, so that there is a smooth crossover
instead of the phase transition \cite{Smi96}.
It is difficult to say  without detailed study
whether large chemical potentials can make the 
transition stronger, but if this happens, it
can have important cosmological   consequences.

\subsection*{Acknowledgments}

I am grateful to N.~Agasian,
R.~Brandenberger, J.~Ng and A.~Zhitnitsky for 
discussions.
This work was supported by 
NSERC of Canada, by Pacific Institute for the Mathematical Sciences
and in part by RFBR grant 
00-15-96557 for the promotion of scientific schools.


\end{document}